\documentclass[twocolumn,amssymb,amsmath,aps,secnumarabic,floatfix]{revtex4}
\usepackage{xspace}
\usepackage{graphicx}

\newcommand{\etal}{\emph{et~al}.\@\xspace}
\newcommand{\ie}{\emph{i.e.}\@\xspace}
\newcommand{\GC}{\mathcal{G}}
\newcommand{\flabel}[1]{\label{fig:#1}}
\newcommand{\elabel}[1]{\label{eq:#1}}
\newcommand{\Fref}[1]{Fig.~(\ref{fig:#1})}
\newcommand{\Eref}[1]{Eq.~(\ref{eq:#1})}
\newcommand{\ave}[1]{\left\langle #1 \right\rangle}
\newcommand{\plaind}{\mathrm{d}}

\expandafter\ifx\csname package@font\endcsname\relax\else
 \expandafter\expandafter
 \expandafter\usepackage
 \expandafter\expandafter
 \expandafter{\csname package@font\endcsname}
\fi
\begin{document}
\title{Comment on: ``Avalanches and Non-Gaussian Fluctuations of the
Global Velocity of Imbibition Fronts''}
\author{Gunnar Pruessner}
\email{g.pruessner@imperial.ac.uk}
\affiliation{
Department of Mathematics, Imperial College London, London SW7 2AZ}

\date{\today}
\maketitle

The theoretical study of interfaces in disordered media has led to many
interesting results, many of which have
never been experimentally verified. The work
\cite{PlanetSantucciOrtin:2009} by Planet, Santucci and
Ort{\'i}n is therefore of great importance. 
However, as shown here,
their data analysis used in \cite{PlanetSantucciOrtin:2009} pre-determines 
the universal exponents to unity. 

The avalanche size distribution in interfacial growth
\cite{PlanetSantucciOrtin:2009} is expected to display simple scaling
in the form
\begin{equation}
P_s(s)=a_s s^{-\alpha} \GC_s(s/s_c)
\elabel{simple_scaling}
\end{equation}
where $P_s(s)$ describes the probability density of finding an avalanche
of size $s$. A corresponding scaling ansatz with exponent $\tau$ is made for the avalanche
duration $T$. 

In \cite{PlanetSantucciOrtin:2009}, to de-dimensionalise the data,
measurements of individual avalanche
sizes were divided by the average,
\ie the histogram $H_u(u)$ was analysed
where $u=s/\ave{s}$, with the aim to extract $\alpha$ as defined in
\Eref{simple_scaling}. The same technique had been employed for
the avalanche duration distribution $P_T(T)$, using the histogram
$H_w(w)$ with $w=T/\ave{T}$.

As shown in the following, the scaling exponents $\alpha$ and 
$\tau$ are actually bound to be unity once the
histograms $H_u(u)$ and $H_w(w)$ display a collapse, as found in
\cite{PlanetSantucciOrtin:2009}. 

The dimensionless histogram $H_u(u)$ is related to $P_s(s)$ via
$H_u(u)\plaind u=P_s(s) \plaind s$ where $\plaind u/\plaind s =
1/\ave{s}$. According to \Eref{simple_scaling}, $\ave{s} = A_s
s_c^{2-\alpha}$ for $\alpha<2$, with constant amplitude $A_s$, so that 
\begin{equation}
H_u(u) = \tilde{A}_s u^{-\alpha} s_c^{(1-\alpha)(2-\alpha)} \GC_s(u A_s s_c^{1-\alpha})
\ .
\elabel{Hu_scaling}
\end{equation}
The \emph{only} way for this function to collapse, i.e. to be
independent from $s_c$, as found by Planet \etal, is that $\alpha=1$ and
by the same argument $\tau=1$. To prevent artefacts,
the \emph{only} scale allowed to de-dimensionalise the observable
(as often required in experimental data analysis) is its cutoff or any
measure thereof, e.g. $u=s/s_c$. To produce a collapse, its suitably
rescaled histogram, for example $H(u)u^{\alpha} s_c^{\alpha-1}$, is 
to be plotted versus the dimensionless quantity $u$.

A few remarks are in order: Firstly, the slope of $H_u(u)$ in a double
logarithmic plot measures the scaling of the product of the power law
prefactor and the scaling function in \Eref{Hu_scaling}, which says little about 
the
actual scaling exponents $\alpha$ and $\tau$ as
defined through \Eref{simple_scaling}. 
This form of an ``apparent exponent'' \cite{ChristensenETAL:2008} is 
what was \emph{actually measured} in
\cite{PlanetSantucciOrtin:2009} in an intermediate region of the
incomplete collapse. 
Secondly, a proper data collapse
is achieved for $P(s)\ave{s}^{\alpha/(2-\alpha)}$ 
or $P(s)s^\alpha$ plotted versus $s/\ave{s}^{1/(2-\alpha)}$. 
\Fref{example} shows attempts to collapse a (mock) dataset known
analytically to have $\alpha=4/3$ along the lines above.

The other scaling feature Planet \etal determine is the joint
histogram $H(u,w)$, suggesting that its centre of mass follows roughly 
$u\propto w^{1.31}$. This behaviour, again, is normally expected for the
dimensionful observables $s$ and $T$
\cite{ChristensenFogedbyJensen:1991}.
If the conditional average $\ave{s|T}$ follows asymptotically
$T^\gamma$, then
$u=\frac{s}{\ave{s}} \propto \frac{T^\gamma}{s_c^{2-\alpha}} \propto
\left(\frac{T}{T_c^{2-\alpha}}\right)^{\gamma}$
where $s_c\propto T_c^\gamma$ was used. The scaling $u\propto
w^x$ found by Planet \etal follows only if $\alpha=\tau$. 

One might
speculate that the process considered by Planet \etal belongs to the C-DP
universality with
$\alpha=1.11(5)$, $\tau=1.17(5)$
\cite{Bonachela:2008} and $x=1.5(2)$ \cite{NakanishiSneppen:1997},
which is possibly revealed by an appropriate data collapse.

\begin{figure}[b]
\includegraphics*[width=0.95\linewidth]{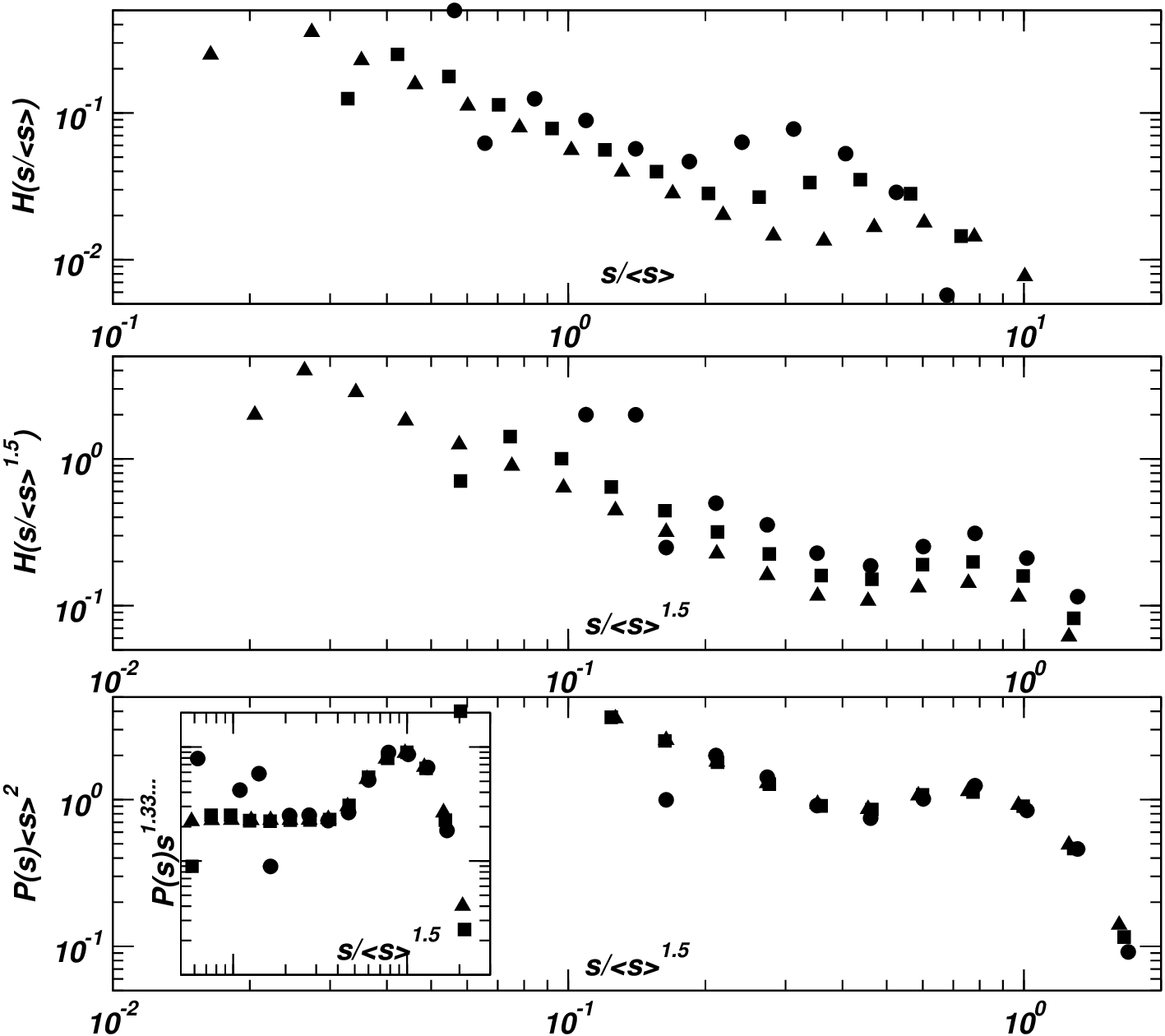}
\caption{\flabel{example}
Attempted collapse of noisy data, known analytically to follow
\Eref{simple_scaling} with $\alpha=4/3$
\cite{Pruessner_exactTAOM:2003}, using
$H_u(u)=P_s(s) \plaind s/\plaind u$ versus $u=s/\ave{s}$ in the top graph, 
$H_u(u)=P_s(s) \plaind s/\plaind u$ versus $u=s/\ave{s}^{1/(2-\alpha)}$ in the middle graph, 
and 
$P_s(s) \ave{s}^{\alpha/(2-\alpha)}$ versus $s/\ave{s}^{1/(2-\alpha)}$ 
in the bottom graph. The inset shows the widely used collapse
$P_s(s)s^{\alpha}$ versus $s/\ave{s}^{1/(2-\alpha)}$, where
$s_c$ is expressed as $\ave{s}^{1/(2-\alpha)}$.
}
\end{figure}

\bibliography{articles,books}

\end{document}